\documentstyle[12pt,epsfig]{article}
\newlength{\dinwidth}
\newlength{\dinmargin}
\newcommand\ben{\begin{equation}}
\newcommand\een{\end{equation}}
\newcommand\bea{\begin{eqnarray}}
\newcommand\eea{\end{eqnarray}}
\newcommand\cD{\cal{D}}

\newcommand\nn{\nonumber}

\newcommand{\bc}{\begin{center}}
\newcommand{\ec}{\end{center}}

\newcommand{\p}{\partial}

\newcommand{\vk}{{\vec{k}}}

\newcommand{\cO}{{\mathcal O}}

\newcommand\vphi{\vec{\phi}}

\begin{document}
\thispagestyle{empty}
\addtocounter{page}{-1}
\vskip-0.35cm
\begin{flushright}
UK/09-04 \\
\end{flushright}
\vspace*{0.2cm}
\centerline{\Large \bf $CP^{N-1}$ Models at a Lifshitz Point} 
\vspace*{2.0cm}
\centerline{\bf 
Sumit R. Das and Ganpathy Murthy}
\vspace*{0.35cm}
\centerline{\it Department of Physics and Astronomy,}
\vspace*{0.2cm}
\centerline{\it University of Kentucky, Lexington, KY 40506 \rm USA ${}^b$}
\centerline{\tt das@pa.uky.edu, murthy@pa.uky.edu}

\vspace*{3.0cm} \centerline{\bf Abstract} \vspace*{0.3cm} We consider
$CP^{N-1}$ models in $d+1$ dimensions around Lifshitz fixed points
with dynamical critical exponent $z$, in the large-N expansion. It is
shown that these models are asymptocially free and dynamically
generate a mass for the $CP^{N-1}$ fields for all $d=z$.  We
demonstrate that, for $z=d=2$, the initially nondynamical gauge field
acquires kinetic terms in a way similar to usual $CP^{N-1}$ models in
$1+1$ dimensions. Lorentz invariance emerges generically in the
low-energy electrodynamics, with a nontrivial dielectric constant
given by the inverse mass gap and a magnetic permeability which has a
logarithmic dependence on scale. At a special multicritical point, the
low-energy electrodynamics also has $z=2$, and an essentially singular
dependence of the effective action on $B=\epsilon_{ij}\partial_iA_j$.

\vspace*{0.5cm}
\baselineskip=18pt

\newpage
Nonlinear sigma models are
ubiquitous in a variety of areas in theoretical physics. In this paper
we will deal with the $CP^{N-1}$ model \cite{D'Adda:1978uc}, 
whose fields are $N$ component
complex vectors $\vphi(t,x)$ constrained by
\ben
\vphi^\star \cdot \vphi = \frac{1}{g^2}
\label{twoa}
\een
and fields which differ by an overall (space-time dependent) phase are
identified,
\ben
\vphi (t,x) \sim e^{i\theta(t,x)}~\vphi (t,x)
\label{twoc}
\een
The identification is incorporated by introducing a non-dynamical $U(1)$ gauge
field, $A_\mu$. The conventional, relativistic,  action is
\ben
S = \frac{1}{2} \int dt\int d^d x~ (D_\mu \vphi)^\star(D^\mu \vphi)
\label{three}
\een
where
\ben
D_\mu \equiv \p_\mu + iA_\mu
\label{four}
\een
Integrating out $A_\mu$ leads to a nonlinear action 
which involves only the $\vphi$ fields. 
As is well known, in $d=1$ the model (\ref{three}) 
is asymptotically free and
generates a mass $m$ for the fields $\vphi$ by dimensional transmutation, 
as can be explicitly seen in the 't Hooft large-N expansion
$N \rightarrow \infty~~~~g \rightarrow 0~~~~~g^2N = \lambda = {\rm
  fixed}$
\cite{D'Adda:1978uc}. At the same time, the initially nondynamical
gauge field acquires a standard kinetic energy term, with a gauge coupling
constant given by $m^2$. This is the simplest example of a dynamical 
emergence of gauge dynamics.

The $d=2$ model is interesting for condensed matter applications.  In
fact, the $O(3)$ nonlinear sigma model with three component unit
vector $\hat{n}$ can be rewritten as the $CP^1$ model via the
identification $\bar{\phi}\vec{\sigma}\phi=\hat{n}$. It is evident
that local phase transformations of the $\phi$ fields do not affect
the ``gauge-invariant'' field $\hat{n}$. Now there is a usual
order-disorder transition: In the magnetically ordered phase of
$\hat{n}$, the $\phi$ field is condensed, and gauge field is gapped
out by the Higgs mechanism.  However, gauge field dynamics appears in
the disordered phase when the $\phi$ fields become massive.  Normally,
the gauge fields also become massive and $\phi$ fields become confined
on the paramagnetic side of the transition due to the compactness of
the gauge field \cite{Polyakov:1975rs}. However, suppressing the
monopoles \cite{lau-dasgupta} of the gauge field (which correspond to
``hedgehog'' configurations of the original $\hat{n}$ fields) leads to
a new critical point \cite{kamal}, and a paramagnetic phase with a
gapless photon \cite{motrunich}. There are conjectures that such a
model with a noncompact gauge field also describes a possible
non-Landau, deconfined, critical point \cite{deconfined-criticality}
between the Ne\'el and bond-ordered phases of the $d=2$ quantum
antiferromagnet.

In this paper we consider UV modifications of these models, 
which correspond to Lifshitz-like
fixed points with a dynamical critical exponent $z$,
\ben
S_L = \frac{1}{2} \int dt\int d^d x~ \left[ (D_0 \vphi)^\star(D^0 \vphi)
+ \alpha (D_i \vphi)^\star(D^i \vphi) + |{\cD}^z \vphi|^2 \right]
\label{five}
\een
where the operator ${\cD}^z$ is a sum of $O(d)$ invariant terms
containing $z$ factors of the spatial covariant derivative $D_i$. For
example, in $z=2$
\ben
|{\cD}^z \phi|^2 \equiv  a~(D_iD_j \vphi)^\star \cdot (D_i D_j \vphi) + b~
(D^2 \vphi)^\star \cdot(D^2\vphi) 
\label{six}
\een
with $a,b\ge0$ being parameters.  
For higher $z$ we would have
many more terms corresponding to various orderings of the $D_i$.

At the fixed point $\alpha=0$, one needs to scale the time and space
coordinates as
\ben
t \rightarrow \gamma^z t~~~~~~~~~~~x \rightarrow \gamma t
\label{eight}
\een
Such fixed points, called Lifshitz fixed points, have a variety of
applications in classical condensed matter systems
\cite{Hornreich:1975zz}. They also have a connection to quantum dimer
models \cite{rk}, which are defined with a ``kinetic'' term which
flips dimers on parallel bonds, and a ``potential'' term which gives
an energy to every flippable plaquette. Finally, there is a constraint
that every lattice site should have one and only one dimer touching
it. Recently, it was realized \cite{shivaji1,ardonne} that neutral
(ungauged) one-component Lifshitz fixed points describe special points
(the Rokhsar-Kivelson (RK) points) of quantum dimer models \cite{rk}
on bipartite lattices, where the field $\phi$ is a height variable
dual to a bond which may or may not contain a dimer
\cite{henley,levitov}. The standard RK point describes the transition
between the smooth and rough phases of the height and is
multicritical, in the sense that more than one parameter needs to be
tuned to attain the fixed point \cite{shivaji1,ardonne}. However, it
is possible to construct models with enough symmetries such that the
fixed point can be obtained as a regular crtical point describing, for
example, the phase transition between two different types of
bond-ordered states in a bilayer honeycomb
lattice\cite{vishwanath}. While the above are examples of $z=d=2$
theories of neutral scalars, examples of $z=2$ gauge theories also
occur in condensed matter, in the description of algebraic spin
liquids in $d=3$ \cite{moessner,hermele} and topological critical
phases in $d=2$\cite{shtengel}.

Note that the action of Eq. (\ref{five}) still contains a single
time-derivative of the gauge fields $A_i$, though it has higher
spatial derivatives. Thus, even though one cannot easily integrate out
the $A_\mu$ to obtain a pure spin model, the gauge field is
non-dynamical to begin with. In the spirit of the renormalization
group, one expects that the model defined by Eq. (\ref{five}) for
$N=2$ is in the universality class of a $z=d=2$ O(3) nonlinear sigma
model.

Recently, Lifshitz-type theories have been suggested as UV
completions of low energy Lorentz invariant theories of gravity and
gauge dynamics \cite{Horava:2008jf}.  This is because theories which
are non-renormalizable at the usual Lorentz-invariant UV fixed points
with $z=1$ can become renormalizable for non-trivial $z$. The idea
that Lorentz symmetry violation can be regarded as UV regulators of
field theories has been around for a while. See \cite{Visser:2009fg}
for a recent discussion and references.  In the same spirit, recently
such Lifshitz fixed points have been proposed as UV completions of
four fermion theories similar to the Nambu-Jona Lasinio model in 3+1
dimensions and discussed as possible candidates for physics at the
weak scale \cite{Anselmi:2009vz}, \cite{Dhar:2009dx}.

We will find that for any choice of ${\cD}^z$ the theory defined by
(\ref{six}) is asymptotically free for all $d=z$ and generates a mass
gap, pretty much as the models in \cite{Dhar:2009dx} - so that the
theory is always in a disordered phase. As a result, 
the gauge field acquire a kinetic term.

This means that the possible emergence of Lorentz invariance at
low energies is a little more non-trivial since this has to happen in
the gauge as well in the scalar sector. For $z=d=2$ we explicitly show
that this indeed happens generically, and discuss the possibilities
for higher $d=z$.

For special choices of ${\cD}^z$ (the $a=0$ multicritical point in our
case) something even more interesting happens: one obtains a $z=2$
electrodynamics with a standard $\vec{E}^2$ term, but the leading term
in $B=\epsilon_{ij}\partial_iA_j$ is $(\nabla B)^2$. This naively
suggests that a constant $B$ costs no energy: however a more careful
calculation reveals that there is a nonanalytic dependence on constant
$B$ of the form $B^{3/2}\exp{(-\pi m/B)}$.  Note that the analytic
terms in our $z=2,\ d=2$ electrodynamics have the same form as the
gauge theory descriptions of algebraic spin liquids in $d=3$
\cite{moessner,shtengel}, but appear to be dual to the gauge
description of the transition between two bond-ordered phases
\cite{vishwanath} or the topological critical phase\cite{shtengel} in
$d=2$ in which it is the $\vec{E}^2$ term which is replaced by
$(\epsilon_{ij}\partial_iE_j)^2$.

\section{Asymptotic Freedom and Dynamical Mass Generation}
\label{sectionone}

We will study the large-N limit of the model (\ref{five}) with the
constraint (\ref{twoa}), using standard techniques \cite{Coleman}.
The coupling $g$ in the model (\ref{five}) becomes dimensionless under
Lifshitz scaling (\ref{eight}) when $d=z$. This may be seen by
dimension counting : $t$ has length dimension $z$, so that the length 
dimensions of $\vphi$ and  $g$ are
\ben
[\vphi] \sim [L]^{\frac{z-d}{2}}~~~~~~[g] \sim [L]^{\frac{d-z}{2}}
\label{8}
\een
Whether the coupling is marginally relevant or marginally irrelevant
at $z=d$ depends on the dynamics. To investigate this we use standard large-N
techniques. Imposing the constraint (\ref{twoa}) by a largrange
multiplier field $\chi (t,x)$ we get the action
\ben
S_L =   \frac{1}{2} \int dt\int d^d x~ \left[ (D_0 \vphi)^\star(D^0 \vphi)
+ \alpha (D_i \vphi)^\star (D^i \vphi) + 
|{\cD}^z \vphi |^2 +  \chi (t,x) \left( |\vphi |^2 (t,x) -\frac{1}{g^2}\right) \right]
\label{ten}
\een
Integrating out the field $\vphi$ we get the effective action
\ben
S_{eff} = N \{ {\rm Tr}~\log~\left[ -D_0^2 +
(-1)^z({\cD}^z)^2 + \chi (t,x) \right] -\frac{1}{2g^2}\int dt d^d
x~\chi(t,x) \}
\label{11}
\een 
At $N=\infty$ the functional integral over $A_\mu (t,x) $ and
$\chi(t,x)$ is dominated by the saddle point of (\ref{11}). We will
assume that the
saddle point is translationally invariant and rotationally symmetric
in the $d$ spatial dimensions with a vanishing gauge field strength.
Thus we may set $\chi (t,x)
= \chi_0$ in the saddle point equation \footnote{In principle there
  could a condensation of the field strength. However we will soon see
  in Section 2.1 that for $d=z=2$ the effective action for a constant
  $B=\epsilon_{ij}\partial_iA_j$ is always larger than the action with
  $B=0$. This rules out condensation of $B$. Our assumption that the
  field strength vanishes at the saddle point is thus justified only 
{\em a posteriori}. We do not have a proof that this continues to
hold for all $d=z$, but this appears to be plausible.}. 
This also means that so far as
the saddle point equation is concerned, all possible terms in
${\cD}^z$ contribute equally 
\ben 
2N \int \frac{dk_0~d^d
k}{(2\pi)^{d+1}} \frac{1}{k_0^2 + \alpha \vk^2 + (\vk^2)^z + \chi_0} =
\frac{1}{g^2}
\label{12}
\een
For any finite $\alpha$, 
this integral is logarithmically divergent for $d=z$ and behaves as
$\log \frac{\Lambda^{2z}}{m^2}$, where $\Lambda$ is a cutoff on
the spatial momentum $\vk$. This immediately implies that a
solution to the gap equation is
\ben
m^2 \equiv \chi_0 \sim \Lambda^{2z} {\rm exp}~[-\frac{A}{g^2N} ]
\label{13}
\een
where $A$ is a {\em positive} real number . Since $m$ is (to leading
order in $1/N$) the physical mass of the $\vphi$ field (i.e. in a
lorentzian signature this is the lowest value of the energy of 
a single particle state), it is clear
from (\ref{13}) that the coupling $g^2$ has to be asymptotically free,
with a beta function
\ben
\Lambda \frac{d}{d\Lambda} g = - \frac{g^3 N}{A}
\een
It is useful to evaluate the integral in (\ref{12}) for our primary case of
interest, $z=d=2$. For $\alpha, m \ll \Lambda^2$ we get
\ben
m =  2 \Lambda^2~e^{-\frac{2 \pi}{g^2N}} -
  \frac{\alpha}{2}
\een
The standard gaussian fixed point corresponds to $\alpha \gg
\Lambda^2$ and leads to a linearly divergent answer in this case.

Dynamical mass generation for this model is thus almost exactly
identical to that in the four-fermion model of Ref. \cite{Dhar:2009dx}.
The effective action for the gauge field $A_\mu$ and the fluctuations
of $\chi(t,x)$ has to be now obatined by substituting
\ben
\chi (t,x) = \chi_0 + \frac{1}{\sqrt{N}} \delta \chi~~~~~~~
A_\mu (t,x) \rightarrow \frac{1}{\sqrt{N}} A_\mu (t,x)
\label{15}
\een
in (\ref{11}). Clearly, this will generate kinetic terms for $\delta
\chi$ and $A_\mu$. The effect of these will be to provide corrections
to the leading order propagator of the $\vphi$ fields which is simply
the integrand of (\ref{12}). Accordingly, the parameter $\alpha$ will
be renormalized. If we go off the critical surface containing the
Lifshitz fixed point, the renormalized value of $\alpha$ will be
nonzero. Clearly, when the spatial momenta are much smaller than
${\sqrt{\alpha}}$, the propagator of the $\vphi$ will be dominated by
the $\alpha_{ren} k^2$ term. Therefore at low energies when
$\alpha\ne0$, Lorentz invariance is recovered with a speed of light
given by $\frac{1}{\sqrt{\alpha}}$.

\section{Effective Action for the Gauge Fields : $d=z=2$, $\alpha=0$}
\label{sectiontwo}

Emergence of Lorentz symmetry at low energies in the gauge field
sector is more non-trivial, especially when $\alpha=0$, which is the
case we will concentrate on. By gauge invariance, the induced action
for the gauge fields must be functionals of the field strengths
$F_{0i}$ and $F_{ij}$ and their derivatives. In addition, it must be
symmetric under spatial rotations.  For a Lorentz symmetry to emerge,
this effective action must contain combinations like
\ben
\epsilon_0 F_{0i}^2 + \frac{1}{\mu_0} F_{ij}^2
\label{17}
\een
with constant $\epsilon_0,\mu_0$. In that one can
now rescale $t,x,A_0,A_i$ to get a standard Lorentz invariant form

The length dimensions of the dielectric constant $\epsilon_0$ and 
magnetic permeability $\mu_0$ may be easily seen to be
\ben
[\epsilon_0] \sim [L]^{z-d+2}~~~~~~~[\mu_0] \sim [L]^{d+z-4}
\label{17a}
\een
so that the speed of light $c = 1/{\sqrt{\mu_0\epsilon_0}}$ 
has length dimensions
\ben
[c] \sim [L]^{1-z}
\een
as it should.

It is not at all obvious that terms like (\ref{17}) {\em have} to
emerge at $\alpha=0$, since the parent theory has $z=2$.  In fact we
will show that for special choices of the operator ${\cD}^z$ this will
{\em not} happen. However, we will find that for generic choices of
${\cD}^z$, terms like (\ref{17}) do appear.

Let us first address this question for $z=d=2$, using the form
(\ref{six}).
For this purpose, it is sufficient to consider the
effective action (\ref{11}) with $\delta \chi = 0$, so that we
essentially have
\ben
S_{eff} = N \{ {\rm Tr}~\log~\left[ -D_0^2 +  
(a+b)(D_iD^i)^2 + a (B(t,x))^2 -ia \epsilon^{ij}(\partial_i
  B)D_j + m^2 \right] - (B=0~~ term)\}
\label{18}
\een
where we have used the commutation relation
\ben
[D_i,D_j] = i F_{ij}
\een
and for $d=2$ renamed $F_{12} = B (t,x)$.

\subsection{Constant Magnetic Field}

It is useful to first evaluate this for a constant $B$. Then the
problem in evaluating the effective action reduces to the problem of
determining the eigenvalues of the operator
\ben
H(B) = -D_0^2 + (a+b) (-D_1^2 - D_2^2)^2 + a B^2
\een
which is closely related to the problem of Landau diamagnetism. Let us
choose a Landau gauge
\ben
A_0 = A_1 = 0~~~~~~~A_2 = B~x^1
\een
Consider the system to be in a large box with size in the time
direction $T$ and spatial sizes $L_1,L_2$. For large enough
$T,L_1,L_2$ the eigenvalue of $\p_0$ can be taken to be continuous,
which we will call $p_0$. 
It is straightforward to see that the eigenvalues of $H(B)$ are
\ben
\kappa(p_0,n) = p_0^2 + (a+b)B^2 (2n+1)^2 + a B^2
\label{22a}
\een
with a degeneracy of the level $n$ given by
\ben
d(n) = \frac{B L_1 L_2}{2\pi}
\label{23}
\een
To evaluate the effective action (\ref{18}) we use the
Nambu-Schwinger-de Witt representation,
\ben
-S_{eff} = N \int_0^\infty \frac{ds}{s}~e^{-m^2 s}~{\rm
  Tr}~e^{-sH(B)} - (B=0~term)
\label{24}
\een
Using (\ref{23}) and (\ref{22a}) we have
\bea
{\rm Tr}~e^{-sH(B)} & = & e^{-s a B^2} \sum_{n=0}^\infty~T
\int_{-\infty}^\infty \frac{dp_0}{2\pi}~
\frac{B L_1 L_2}{2\pi}~
e^{- [ sp_0^2 + 4(a+b)B^2 s (n+\frac{1}{2})^2 ] } \nn \\
& = & \frac{VT~B}{8\pi^2}\sqrt{\frac{\pi}{s}}e^{-saB^2}
~{\Large \vartheta_2}~ [ 0~ | ~4iB^2 s (a+b)/\pi ] \nn \\ 
\label{25}
\eea
where $V=L_1L_2$ denotes the spatial volume and $\vartheta_2[w|\tau]$ 
is a Jacobi theta function \footnote{We are grateful to Al Shapere
for pointing out that an efficient way to manipulate this sum is to 
recognize this as a theta function.}.
To examine the small $B$ behavior it is useful to use standard theta
function identities to write
\bea
{\rm Tr}~e^{-sH(B)} & = & 
\frac{VT}{16 \pi s~\sqrt{a+b}}e^{-saB^2}
~\vartheta_4~ [ 0 ~|~ i\pi/(4B^2 s (a+b)) ] \nn \\
& = &  \frac{VT}{16 \pi s~\sqrt{a+b}}e^{-saB^2}
\sum_{k=-\infty}^{\infty} (-1)^k~e^{-\frac{\pi^2 k^2}{4sB^2(a+b)}}
\label{25a}
\eea 
The theta function in (\ref{25a}) may be written in a product
represenation as 
\ben
\vartheta_4~[0~|~ i\pi/(4B^2 s (a+b)) ] = \prod_{n=1}^\infty
( 1 - e^{-\frac{(2n-1)\pi^2}{4B^2 s (a+b)}})^2~(
1- e^{-\frac{2n\pi^2}{4B^2 s (a+b)}})
\een
Since $a,b \geq 0$, this immediately shows that
\ben
{\rm Tr}~e^{-sH(B)} < {\rm Tr}~e^{-sH(0)}
\een
so that
\ben
S_{eff}(B) > S_{eff}(0)
\label{25z}
\een
for any $m^2$. This provides a justification for setting $B=0$ in the
saddle point equation which determines $m^2$.
The result (\ref{25z}) in fact holds for all $d$ with
$z=2$.
 
The integral over $s$ in (\ref{25a}) 
can be performed, leading to the effective action 
\footnote{Note that there is an overall factor of $N$ in the effective action
(\ref{36a}). Since we are performing a $1/N$ expansion, so that the
fields have to be rescaled as in (\ref{15}). The factor of $N$ cancels
for the terms which are quadratic in the fields.}
\ben 
\frac{S_{eff}(B)}{VT} =
-\sum_{k \neq 0} (-1)^k \frac{ Bm}{4\pi^2 k}
\sqrt{1+\frac{aB^2}{m^2}}~
K_1 \left( \frac{\pi k m}{B}\sqrt{\frac{1+\frac{aB^2}{m^2}}{a+b}} \right)
- \frac{1}{16\pi \sqrt{a+b}}\int_0^\infty \frac{ds}{s^2}e^{-m^2 s}
\left( e^{-s a B^2} -1 \right)
\label{25c}
\een
where $K_1$ denotes a modified Bessel function.
In deriving (\ref{25c}) 
we have noted that the $k=0$ term in the sum in (\ref{25a}) is
the sole contribution when $B=0$ and subtracted that. 

The first term on the right hand side of (\ref{25c}) has a non-analytic
dependence on $B$ for small $B$.
This follows from the asymptotic behavior of the modified
Bessel function.  For $B \ll m$ the sum in (\ref{25c}) is 
dominated by the $k=1$
term, which leads to
\ben
\frac{S_{eff}(B)-S_{eff}(0)}{VT}\simeq 
\frac{B^{\frac{3}{2}}~m^{\frac{1}{2}}~b^{\frac{1}{4}}}{4\pi^2\sqrt{2}}
~e^{-\frac{\pi m}{B\sqrt{b}}}
\label{31a}
\een
The second term contains various powers of $B$.

Therefore, at the multicritical
point $a=0$, the effective action vanishes for $B=0$ in a {\em
non-analytic fashion} \footnote{We have performed the sum in
(\ref{25a}) using a Euler-McLaurin expansion and verified that there
are no polynomial terms in $B$ in this case to very high
orders.}.

For any $a \neq 0$ the 
final form of the effective action begins with a term 
proportional to $B^2$,
It follows from
(\ref{24}) that the coefficient of this term is divergent
(proportional to $\Gamma(0)$). To understand this, we use a version of
dimensional regularization by adding $(d-2)$ extra spatial directions.
Now the first line of (\ref{25}) will be modified to
\ben
{\rm Tr}~e^{-sH(B)}  =  e^{-saB^2} \sum_{n=0}^\infty~T
\int_{-\infty}^\infty \frac{dp_0}{2\pi}~L^{d-2} \int
\frac{d^{d-2}p}{(2\pi)^{d-2}} 
\frac{B L_1 L_2}{2\pi}~
e^{-sp_0^2 -sp^4 -4(a+b)B^2 s (n+ \frac{1}{2})^2}
\een
This leads to the following coefficient of $B^2$ in the effective action
(\ref{24}) 
\ben 
a
\int_0^\infty \frac{ds}{s} e^{-m^2s}~s^{-\frac{d-2}{4}}
= a\frac{\Gamma(\frac{2-d}{4})}{m^{\frac{2-d}{2}}} 
\een
The coefficient has length dimension $(2-d)$ as required.
In a dimensional regularization scheme we introduce 
a scale $\kappa$ with length dimensions
$-2$, and write this coefficient as
\ben
\frac{1}{\mu_0~\kappa^{\frac{2-d}{2}}}
\label{32}
\een
where $\mu_0$ is dimensionless.
Then as $\epsilon \rightarrow 0$, the finite part of $1/\mu_0$ becomes
\ben
\frac{1}{\mu_0} \sim \log \left( \frac{\kappa}{m} \right)
\een
\subsection{General Calculation Using Heat Kernel}

In
this section we will evaluate the effective action using a standard
heat kernel method for general fields $F_{0i}(t,x)$ and $F_{ij}(t,x)$.
The calculation presented above for constant magnetic field shows that
at the special point $a=0$ the effective action has a non-analytic
dependence on $B$. We would like to determine whether the action
contains terms which are analytic in derivatives of $B$.
As shown in the Appendix , the small $s$ expansion of
the heat kernel is of the form
\ben
{\rm Tr}e^{-sH(E,B)} = \int d^2x~dt \sum_{n=1}^\infty b_n
(x)~s^{\frac{n}{4}-1} 
\label{35a}
\een
which leads to the effective action
\ben
S_{eff} =  N \sum_{n=1}^\infty \int d^2x~dt~b_n(x) 
\frac{\Gamma(\frac{n}{4}-1)}{m^{2(\frac{n}{4}-1)}}
\label{36}
\een
As in the previous subsection the term with $n=4$ can be handled
via dimensional regularization. As explained in the appendix, we should
therefore replace (\ref{36}) by
\ben
S_{eff} =  N \sum_{n=1}^\infty \int d^2x~dt~b_n(x) 
\frac{\Gamma(\frac{n}{4}-1+\frac{\epsilon}{4})}{m^{2(\frac{n}{4}-1+
\frac{\epsilon}{4})}}
\label{36a}
\een
Let us first evaluate this for $a=0$ and $b=1$
For this case, explicit calculations yield 
$b_1(t,x)=b_2(t,x)=b_3(t,x)=b_4(t,x)= b_5(t,x) = 0$. 
The leading contribution comes from $b_6(t,x)$. After a fairly long
calculation we find that this leads to the effective action
\ben
S_{eff, a=0} = \frac{1}{12m} \int dt d^2x~\left[ F_{0i}^2 +
  \frac{1}{10}(\partial_i B)(\partial^i B) \right]
\een
Higher powers of 
field strength are suppressed by powers of $N$.
Note that the term we obtained in the previous section for constant
$B$ is nonanalytic in $B$, and formally irrelevant by power counting
since its Taylor series is 0. However, in the absence of this
nonanalytic term, any constant $B$ costs no energy, leading to a huge
ground state degeneracy. Thus, the term $B^{3/2}e^{-\pi m/B}$ is a
dangerously irrelevant operator for this special case $a=0,\ \ \ b=1$.
Clearly Lorentz invariance is not regained at low energies in this
case. 

Away from this multicritical point $a \neq 0$. As expected from our
constant $B$ calculation we now find that $b_4 \neq 0$. It is
straightforward to see that
\ben
b_4 (t,x)  \sim  F^{ij} F_{ij}
\een
Since we are interested in the action at low energies, we should
retain only the lowest non-trivial terms with the least number of
derivatives. This leads to the low energy effective action for gauge
fields, upto numerical factors
\ben
S_{eff} \sim \int d^2x~dt~[ \frac{1}{m} F_{0i}F^{0i} + 
\frac{\Gamma(\frac{\epsilon}{4})}{m^{\frac{\epsilon}{2}}} F_{ij}
F^{ij} + \cdots ]
\een
where the ellipses now stand for terms containing more derivatives
and/or more powers of the field strength. As in the previous
subsection, in the spirit of dimensional
regularization this is really
\ben
S_{eff} \sim \int d^2x~dt~[ \frac{1}{m} F_{0i}F^{0i} + 
\log \left( \frac{\kappa}{m} \right)
F_{ij}
F^{ij} + \cdots ]
\een
so that we have $\mu_0$ given in (\ref{32}) and
\ben
\epsilon_0 \sim \frac{1}{m}
\een
Since $\epsilon_0, \mu_0$ are constants independent of $(t,x)$ one can
now rescale $t,x,A_0,A_i$ to get the form
\ben
S_{eff} \sim \int d^2x~dt~[ F_{\mu\nu}F^{\mu\nu}+ \cdots ]
\een
This demonstrates the emergence of approximate low energy Lorentz
symmetry with a scale dependent speed of light. Note that this happens
even when the (renormalized) 
parameter $\alpha_{ren}$ in (\ref{five}) is tuned to zero. At
this Lifshitz point there is no lorentz symmetry in the scalar sector.
For $\alpha_{ren} \neq 0$ the speed of light in the scalar sector is
different from that in the gauge sector.

\section{Discussion}

It is clear from section (\ref{sectionone}) that the coupling $g$ in the
sigma model is asymtptically free for all $d=z$. Does this also mean that
gauge dynamics emerges in all dimensions? From equation (\ref{17a})
we see that the length dimensions of $\epsilon_0$ is always 2 for all
$z=d$. If such a term appears, one would expect that $\epsilon_0 \sim 
m^{2/z}$, since from (\ref{13}) the length dimension of the mass $m$ is
$z$. For $z \geq 3$ it is not clear how such a term can arise 
in the effective action obtained by
integrating out the massive field $\vphi$. In fact if such an
effective action is an expansion of powers of $1/m$ (apart from logs) 
one would expect
that the lowest dimension operator which would appear must have length
dimensions $[L]^{z+d}$. In a similar vein, the magnetic permeability
would have {\em positive} length dimensions. Such a term is also
unlikely to come from an effective action. It would be interesting to
investigate this issue further.

The model studied in this paper gauges the overall $U(1)$ of the
symmetry group. One could, instead, consider gauging the
entire $U(N)$ group to obtain a non-abelian gauged sigma model. This
model would generate a mass gap in exactly the same fashion - in fact
the gap equation is identical. It would be interesting to see the
effective action for the non-abelian gauge fields in this case. It
appears to us that the heat kernel expansion calculation is quite
similar to ours.

Another interesting direction is to revisit the {\em linear} rather
than the non-linear sigma model around a Lifshitz fixed point as has
been originally considered in classical statistical mechanics.
The length dimension of a $(\vphi \cdot \vphi)^2$ coupling is given by
$(d-3z)$, so that this is a relevant operator for $z > d/3$. It would
be interesting to explore if there are IR fixed points for $d > 2$ 
at finite
values of the corresponding coupling, similar to $z=1, d=2$. This could have
interesting applications to particle physics.
These vector models can be also interesting from the point of view of
AdS/CFT correspondence. Lifshitz fixed points have been argued to have
dual gravity descriptions \cite{Kachru:2008yh}. On the other hand, the
dual of usual vector models are higher spin gauge fields in usual
$AdS$ \cite{Klebanov:2002ja}, \cite{Das:2003vw}. 
It would be interesting to see the
nature of the gravity duals for these Lifshitz sigma
models. 

These issues are currently under investigation.

\section{Acknolwedgements} We would like to thank Al Shapere for
enlightening discussions, and the referee for comments about
a previous version of the manuscript.
The work of S.R.D. was supported in part by
a National Science Foundation Grant NSF-PHY-0555444, and that of GM by
NSF-DMR-0703992. GM also thanks the Aspen Center for Physics, where
some of this work was carried out. 

\appendix

\section{Expansion of the Heat Kernel}

In this appendix calculate the heat kernel using the technique of
\cite{Nepomechie:1984wt}. This uses the representation
\ben
{\rm Tr} e^{-s\cO} = \int dt d^2x \int \frac{d\omega
  d^2k}{(2\pi)^3}~e^{-i(\omega t + k \cdot x)}~e^{-s\cO}~
~e^{i(\omega t + k \cdot x)}
\label{35}
\een
Using the basic identity
\ben
e^{-i(\omega t + k \cdot x)}~D_\mu
~e^{i(\omega t + k \cdot x)} = ik_\mu + D_\mu
\een
we have for our case
\ben
e^{-i(\omega t + k \cdot x)}~H(E,B)
~e^{i(\omega t + k \cdot x)}  =  \omega^2 + \vk^4 +a_1 + a_2 + a_3
+a_4
\label{37}
\een
where
\bea
a_1 & = & - 4i k^2~(k^iD_i) \nn \\
a_2 & = & - 2i\omega D_0 - 4 (k^iD_i)^2 - 2 k^2~(D^iD_i) \nn \\
a_3 & = & i [ (D^iD_i)(k^jD_j) +  (k^jD_j)(D^iD_i) + 2 D_i (k^jD_j)
  D^i ] \nn \\
a_4 & = & -D_0^2 + D_iD_jD^iD^j
\eea
Using (\ref{37}) in (\ref{35}) and rescaling
\ben
\omega \rightarrow \frac{1}{s^{1/2}} \omega~~~~~~~~
\vk \rightarrow \frac{1}{s^{1/4}} \vk
\label{46}
\een 
we get
\ben
{\rm Tr}e^{-s H(E,B)} = \frac{1}{s} \int dt d^2x \int \frac{d\omega
  d^2k}{(2\pi)^3}~e^{-(\omega^2 + k^4)}~e^{-G(E,B)}
\label{47}
\een
where
\ben
G \equiv  s^{\frac{1}{4}}a_1 + s^{\frac{1}{2}}a_2 + 
s^{\frac{3}{4}}a_3 + s~a_4
\label{48}
\een
The integrals over $\omega$ and $\vk$ can be now evaluated, leading to 
small-s expansion of the heat kernel of the
form (\ref{35a}), leading to the form of the effective action
(\ref{36}). 

The term with $n=4$ has to be treated in dimensional
regularization. This means that in (\ref{35}) we replace
$d^2k \rightarrow d^d k$, so that after the rescalings (\ref{46}), the
equation (\ref{47}) becomes
\ben
{\rm Tr}e^{-s H(E,B)} = \frac{1}{s^{\frac{d+2}{4}}} 
\int dt d^2x \int \frac{d\omega
  d^dk}{(2\pi)^3}~e^{-(\omega^2 + k^4)}~e^{-G(E,B)}
\label{47a}
\een
For $d = 2 -\epsilon$ we will still evaluate the integrals over
$\vk$ by replacing $d^d k \rightarrow d^2k$ in (\ref{47a}). This leads
to the small $s$ expansion
\ben
{\rm Tr}e^{-sH(E,B)} = \int d^2x~dt \sum_{n=1}^\infty b_n
(x)~s^{\frac{n}{4}-1+\frac{\epsilon}{4}} 
\label{35b}
\een
This leads to the expression (\ref{36a}).

The integrals over $\omega$ and $\vk$ can be performed using basic
symmetry properties. Thus, e.g.
\bea
\int_{-\infty}^\infty 
d\omega~e^{-\omega^2}~\omega^{2n+1} & = &  0~~~~~~~(n~{\rm
  integer}) \nn \\
\int_{-\infty}^\infty 
d\omega~e^{-\omega^2}~\omega^2 & = &  \frac{1}{2} 
\int_{-\infty}^\infty 
d\omega~e^{-\omega^2} \nn \\
\int_{-\infty}^\infty d\omega~e^{-\omega^2}~\omega^4 & = &  \frac{3}{4} 
\int_{-\infty}^\infty 
d\omega~e^{-\omega^2} \nn \\
\cdots~~~~~\cdots
\label{52}
\eea
while
\bea
\int d^2k~k^{i_1}k^{i_2}~~~~k^{1_{2n+1}}~e^{-k^4} & = & 0 ~~~~~~~(n~{\rm
  integer}) \nn \\
\int d^2k~k^{i}k^{j}~~e^{-k^4} & = & \frac{1}{2\sqrt{\pi}} \delta^{ij} 
\int d^2k~e^{-k^4} \nn \\
\int d^2k~(\vk \cdot \vk)^2~~e^{-k^4} & = & \frac{1}{2} 
\int d^2k~e^{-k^4} \nn \\
\int d^2k~(\vk \cdot \vk)^3~~e^{-k^4} & = & \frac{1}{\sqrt{\pi}} 
\int d^2k~e^{-k^4} \nn \\
\int d^2k~(\vk \cdot \vk)^4~k^ik^j~e^{-k^4} & = & \frac{1}{2\sqrt{\pi}} 
\delta^{ij}\int d^2k~e^{-k^4} \nn \\
\cdots~~~~~~~\cdots
\label{53}
\eea
Using these integrals, it is straightforward to see that terms with
$n=1,3,5$ in the sum (\ref{35b}) vanish since they have odd numbers of
$\omega$ and/or $\vk$. A short calculation using the explicit
expressions in (\ref{52}) and (\ref{53}) shows that the various terms
cancel, leading to $b_2 (x,t) = 0$. The first non-trivial term is
therefore for $n=4$. Here, after several cancellations one is left
with 
\ben
b_4 (t,x) \sim F_{ij}F^{ij}
\een
which basically comes from rewriting the second term in $a_4$ as
\ben
D_i D_j D^i D^j = D_i D^2 D_i + D_i D_j [D^i , D^j] 
=  D_i D^2 D_i + \frac{1}{2} [D_i , D_j] [D^i , D^j]
=  D_i D^2 D_i +\frac{1}{2} F_{ij}F^{ij}
\een
The next nonzero term comes at $n=6$. This leads to an electric field
term, which arises from
\ben
\int \frac{d\omega
  d^dk}{(2\pi)^3}~e^{-(\omega^2 + k^4)}~(a_1 a_2 a_1 a_2)
\een
which clearly includes a term
\ben
D_i D_0 D_i D_0
\een
and hence to $F_{0i}F^{0i}$. The $n=6$ term contains other
contributions as well. These contain higher derivative terms in the
field strength $B$. Specifically, for the case $a=0,\ \ b=1$, we
obtain the term
\ben
\frac{1}{120 m}(\nabla B)^2
\een

\end{document}